\documentclass[epj,nopacs]{svjour}
\renewcommand{\bar}[1]{\overline{#1}}
\providecommand{\Journal}[4] {#1 {\bf#2}, #3 (#4)}
\providecommand{\PLB}{Phys. Lett. B} %
\providecommand{\PRL}{Phys. Rev. Lett.} %
\providecommand{\PRD}{Phys. Rev. D}
\providecommand{\NPB}{Nucl. Phys. B} %
\providecommand{\EPJC}{Eur. Phys. J. C} %
\providecommand{\ZPC}{Z. Phys. C} %
\providecommand{\PR}{Phys. Rep.} %
\providecommand{\RMP}{Rev. Mod. Phys.} %

\usepackage{graphics}
\begin{document}
\title{Probing nucleon strange asymmetry
       from charm production
       in neutrino deep inelastic scattering}

\author{Puze Gao \inst{1} \and Bo-Qiang Ma\inst{2,1,3,}\thanks{Email address: mabq@phy.pku.edu.cn; corresponding author. }}

%
%
\institute{Department of Physics, Peking University, Beijing
100871, China  \and CCAST (World Laboratory), P.O.~Box 8730,
Beijing 100080, China \and Di.S.T.A., Universit\`a del Piemonte
Orientale ``A. Avogadro'' and INFN, Gruppo Collegato di
Alessandria, 15100 Alessandria, Italy}

\date{Received: date / Revised version: date}
%
\abstract{We propose a means to detect the nucleon strange
quark-antiquark asymmetry, which is predicted as a
non-perturbative effect, but still unchecked directly by available
experiments. The difference for the $D(c\bar{q})$ and
$\bar{D}(\bar{c}q)$ meson production cross sections in neutrino
and antineutrino induced charged current deep inelastic scattering
is illustrated to be sensitive to the nucleon strange asymmetry.
Prospect is given and the effect due to the light quark
fragmentation is also discussed for the extraction of the strange
asymmetry in future experiments.
\PACS{
        {14.60.Pq}{}  \and
        {12.15.Ff}{}
     } 
} 

\titlerunning{Probing nucleon strange asymmetry
       from charm production
       in neutrino DIS}
\maketitle


\section{Introduction}
\label{intro}

Nucleon structure is a natural laboratory to understand QCD and is
worth to study for its own sake. The nucleon strange
quark-antiquark asymmetry is an interesting feature predicted as a
natural consequence of the non-perturbative aspect of the
nucleon~\cite{bm96,sig87,bur92}. Recently, the nucleon strange
asymmetry has been
suggested~\cite{oln03,kre04,dm04,alw04,dxm04,wak04} as a promising
mechanism to explain the NuTeV anomaly~\cite{NT1,NT2} within the
framework of Standard Model.

While the experimental evidence for such an asymmetry is still
inconclusive, there are some approaches such as the global
analysis of deep inelastic scattering (DIS)
data~\cite{oln03,bar00}, which show a favor for an asymmetric
strange sea, in agreement qualitatively with the intrinsic sea
theory. On the other hand, the CCFR next-to-leading-order (NLO)
analysis of the neutrino induced dimuon production result favors a
symmetric strange sea~\cite{CCFR95}, which is also the case of a
recent NuTeV analysis~\cite{NTmas04}. It seems that more precise
and dedicated research is needed to address the problem in a clear
way.

The measurement of the strange quark distribution relies on
charged current (CC) DIS processes. One method is through parity
violating structure functions for isoscalar target in CC DIS:
$F_3^\nu -F_3^{\bar{\nu}}=2[s(x)+{\bar s}(x)-c(x)-{\bar c}(x)]$,
which gives the total distribution of the strange sea. Another way
is through the combination of CC parity conserving structure
function $F_2^\nu$ with the charged lepton DIS structure function
$F_2^\mu$, for isoscalar target, ${5\over
6}F_2^\nu-3F_2^\mu=x[{4\over 3}s(x)-{1\over 3}{\bar s}(x)-c(x)]$.
Such an idea has been applied using high statistic neutrino and
charged lepton nucleon DIS data, and the result at low $x$ shows a
sizable disagreement with the direct measurement from the CCFR
dimuon result~\cite{con98}. The extraction of a small quantity
from the difference of two large quantities may suffer from
systematic uncertainties, which also seems to be the case for the
extraction of the strange asymmetry by CC parity conserving
structure functions: $F_2^\nu-F_2^{\bar\nu}=2x[s(x)-{\bar s}(x)]$.

A method free from the above drawback is to use the charged
current charm production process, which is the main idea of the
CCFR and NuTeV dimuon experiments~\cite{CCFR95,CCFR93,NTdimu01},
with its leading order (LO) subprocesses being $\nu_\mu s
\rightarrow \mu^- c$ and $\nu_\mu d \rightarrow \mu^-  c$. The
latter subprocess is Cabibbo suppressed, thus the charm production
in $\nu$-induced process is most sensitive to the strange quark
distribution in the nucleon. Similarly, the anticharm production
in $\bar\nu$-induced process is sensitive to the antistrange quark
distribution, as the corresponding partner subprocesses are
$\bar{\nu}_\mu \bar{s} \rightarrow {\mu}^+  \bar{c}$ and
$\bar{\nu}_\mu \bar{d} \rightarrow {\mu}^+ \bar{c}$, with the
latter subprocess being Cabibbo suppressed.

The oppositely charged dimuon signature is easy to identify and
measure in massive detectors, which allow for the collection of
high statistics data samples, {\it e.g.}, the CCFR experiment has
a sample of data with 5030 $\nu_\mu$ induced events and 1060
$\bar\nu_\mu$ induced events, and the NuTeV has 5102 $\nu_\mu$
induced events and 1458 $\bar\nu_\mu$ induced
events~\cite{NTdimu01}. However, these two experiments neither
show strong support for an asymmetric strange sea, nor can they
rule it out~\cite{bm96,kre04,bar00}. There are uncertainties in
the estimation of the semi-muonic decay of the charmed
hadrons~\cite{CCFR95,CHOR02}, {\it e.g.}, the average
semi-leptonic branching ratio for $\nu$ and $\bar\nu$ induced
events was only constrained by ${{{\bar B}_c-{\bar B}_{\bar
c}}\over {\bar B}_c}={{0.011\pm0.011}\over 0.1147}\sim
0-20\%$~\cite{CCFR95}. Besides, the interplay of strange asymmetry
and the light quark fragmentation (LQF) effect, as will be
discussed in section~4, can only be drawn more clearly in
inclusive measurement of charged and neutral charm productions.
Thus a direct measurement of charmed hadrons produced in $\nu$ and
$\bar\nu$ induced CC DIS will provide more valuable information to
probe the s and $\bar s$ distributions of the nucleon. It is the
purpose of this work to show that inclusive charm productions in
neutrino and antineutrino induced CC DIS processes will be a
promising way to detect the strange quark-antiquark asymmetry.


\section{Charged current charm production}

The differential cross section for charmed hadron $H^+$ production
in neutrino induced CC DIS can be factorized as
\begin{eqnarray}
{d^3\sigma_{\nu_{\mu}N\rightarrow \mu^-H^+X}\over d\xi dy dz}
=\sum_q{d^2\sigma_{\nu_{\mu}N\rightarrow \mu^- q X}\over d\xi dy}
D_q^{H^+}(z)\;,
\end{eqnarray}
where the function $D_q^{H^+}(z)$ describes the fragmentation of a
quark q into the charmed hadron $H^+$, with $z$ being the momentum
fraction of the quark $q$ carried by the produced hadron $H^+$.
For the purpose of this article, the charmed hadron $H^+$ is taken
to be $D^+(c\bar d)$ or $D^0(c\bar u)$ meson, with $H^-$ denoting
its antiparticle $D^-({\bar c} d)$ or ${\bar D^0}({\bar c} u)$.

It is generally believed that the possibility for light quark
fragmentation into charmed hadrons is very small. For example, the
Lund string model implemented in some popular Monte Carlo programs
predicts a suppression proportional to $\exp(-bm_q^2)$ for $q\bar
q$ production in the process of hadronization~\cite{string81}.
With a knowledge of the strange suppression $\lambda\sim
0.3$~\cite{laf95,SLD97}, the suppression for charm will be lower
than $10^{-4}$, which can be safely neglected.

In this case, at leading order, only the $\nu_\mu s\rightarrow
c\mu^-$ and $\nu_\mu d\rightarrow c\mu^-$ subprocesses contribute
to charmed hadron production. For isoscalar target and neglecting
target mass effects, the leading order differential cross section
for charm production is given by~\cite{CCFR95,CCFR93}:
\begin{eqnarray}
{d^2\sigma_{\nu_{\mu}N\rightarrow \mu^- c X}\over d\xi dy}
&=&{2G^2ME_\nu\over \pi(1+Q^2/M_W^2)^2}\left(1-{m_c^2\over
2ME_\nu\xi}\right) \nonumber
\\& \times&
\xi\left[{d(\xi)+u(\xi)\over 2}|V_{cd}|^2 +
s(\xi)|V_{cs}|^2\right] ,\label{nucharm}
\end{eqnarray}
where $\xi$ is the momentum fraction of the struck quark in the
infinite momentum frame. It is introduced with the consideration
of a non-negligible charm quark mass, and is related to Bjorken
scaling variable $x$ through (neglecting light quark mass):
$\xi\approx x(1+m_c^2/Q^2)$, referred to as slow-rescaling. The
term $(1-m_c^2/2ME_\nu\xi)$ in Eq.~(\ref{nucharm}) is introduced
as an energy threshold for charm production and is supported by
experiments~\cite{NOM00}.

\section{Probing the nucleon strange asymmetry}

The differential cross sections for charmed hadrons, namely, $H^+$
($D^+$ or $D^0$) and $H^-$ ($D^-$ or $\bar D^0$), produced in
$\nu$ and $\bar\nu$ induced CC DIS respectively, are closely
related to the $s$ and $\bar s$ distributions of the nucleon, and
their difference, as can be seen in the following, is quite
sensitive to the nucleon strange asymmetry.

Neglecting the light quark fragmentation effect, and using
Eq.~(\ref{nucharm}) and its corresponding partner process for
${\bar c}$ production ${\bar\nu}_\mu N \rightarrow {\bar
c}\mu^+X$, we can write the difference between $H^+$ and $H^-$
production cross sections in CC DIS:
\begin{eqnarray}
f_{H^+}-f_{H^-}&\equiv& {d^3\sigma_{\nu_{\mu}N\rightarrow \mu^-
H^+ X}\over d\xi dy dz}- {d^3\sigma_{\bar\nu_{\mu}N\rightarrow
\mu^+ H^- X}\over d\xi dy dz} \nonumber \\ &=&{2G^2ME_\nu\over
\pi(1+Q^2/M_W^2)^2}\left(1-{m_c^2\over 2ME_\nu\xi}\right)
\nonumber
\\ &\times&
\bigg{\{}{1\over 2}\xi[d_v(\xi)+u_v(\xi)]|V_{cd}|^2
\nonumber\\
&& ~+ \xi[s(\xi)-\bar
s(\xi)]|V_{cs}|^2\bigg{\}}D_c^{H^+}(z)\;,\label{csd}
\end{eqnarray}
where charge symmetry $D_c^{H^+}(z)=D_{\bar c}^{H^-}(z)$ for
fragmentation process is assumed, and $u_v(\xi)\equiv u(\xi)-{\bar
u}(\xi)$ and $d_v(\xi)\equiv d(\xi)-{\bar d}(\xi)$ are valence
quark distributions of the proton.

From Eq.~(\ref{csd}), one sees that two terms, ${1\over
2}\xi[d_v(\xi)+u_v(\xi)]$ and $\xi[s(\xi)-\bar s(\xi)]$,
contribute to the cross section difference $f_{H^+}-f_{H^-}$, with
$|V_{cd}|^2\simeq 0.05$ and $|V_{cs}|^2\simeq 0.95$~\cite{PDG04}
being their respective weights. The strange asymmetric part of
Eq.~(\ref{csd}) can be estimated from an integral on variable
$\xi$, to contribute a fraction
\begin{eqnarray}
P_{\mathrm{SA}}\approx{2S^-|V_{cs}|^2\over
Q_V|V_{cd}|^2+2S^-|V_{cs}|^2}\;,
\end{eqnarray}
to the integral of the cross section difference $\int
d\xi(f_{H^+}-f_{H^-})$. Here, $S^-$ and $Q_V$ are defined as
$S^-\equiv\int \xi[s(\xi)-\bar
 s(\xi)]d\xi$ and $Q_V\equiv \int  \xi[d_v(\xi)+u_v(\xi)]d\xi$.

In Table~1, results of the strange asymmetry from some models
accounting for the NuTeV anomaly are listed, together with our
estimations of the contributions due to strange asymmetry to the
the cross section difference $f_{H^+}-f_{H^-}$, namely, the $\xi$
integrated fraction $P_{\mathrm{SA}}$.

\begin{table*}
\caption{Contributions of s/${\bar s}$ asymmetry to NuTeV anomaly
and to $f_{H^+}-f_{H^-}$}
\label{tab:1}       
\begin{tabular}{lllll}
\hline\noalign{\smallskip}
 Models & $Q^2$ &
To NuTeV anomaly & $2S^-/Q_V$ & To $f_{H^+}-f_{H^-}$  \\
\noalign{\smallskip}\hline\noalign{\smallskip}
Ding-Ma\cite{dm04} & $Q_0^2$ &
 $30\%\sim 80\%$ & $0.007\sim 0.018$ & $12\%\sim 26\%$ \\
Alwall-Ingelman~\cite{alw04} & 20~GeV$^2$& $30\%$ & $0.009$ &
 $15\%$ \\
 Ding-Xu-Ma \cite{dxm04} & $Q_0^2$ &
 $60\%\sim 100\%$ & $0.014\sim 0.022$ & $21\%\sim 29\%$ \\
Wakamatsu~\cite{wak04} & 16~GeV$^2$& $70\%\sim 110\%$ & $0.022\sim
0.035$ &
$30\%\sim 40\%$ \\
\noalign{\smallskip}\hline
\end{tabular}
\end{table*}

As shown in Table~1, from the model
calculations~\cite{dm04,alw04,dxm04,wak04} that can explain the
NuTeV anomaly, the strange asymmetry contributes a sizable
proportion $(12\%\sim 40\%)$ to the cross section difference. Note
that the distribution functions $\xi[d_v(\xi)+u_v(\xi)]$ and
$\xi[s(\xi)-\bar s(\xi)]$ may evolve with $Q^2$, turning flatter
and shifting towards smaller $\xi$ region as $Q^2$ increases.
However, their relative feature will remain and the proportion of
$S^-$ to $Q_V$, will be of the same order in larger $Q^2$ and in
$Q_0^2$. Thus, as to their relative feature, it does not matter
much whether the parton distributions are taken at $Q_0^2$ or at
larger $Q^2$. Since the peak of $\xi[s(\xi)-\bar s(\xi)]$ is
confined in narrower $\xi$ region than $\xi[d_v(\xi)+u_v(\xi)]$,
its contribution is expected to be more prominent than the
integrated one in Table~1. Thus it is promising to measure the
strange quark-antiquark asymmetry from $f_{H^+}-f_{H^-}$.

Compared to the sum of the cross sections $f_{H^+}+f_{H^-}$, the
cross section difference $f_{H^+}-f_{H^-}$ is not a very small
quantity, as can be seen from the ratio of their integrals,
\begin{eqnarray}
R\equiv{\int d\xi(f_{H^+}-f_{H^-})\over\int d\xi(f_{H^+}+f_{H^-})}
\approx {Q_V|V_{cd}|^2+2S^-|V_{cs}|^2\over
(Q_V+2Q_S)|V_{cd}|^2+2S^+|V_{cs}|^2}\;,
\end{eqnarray}
where
 $Q_S\equiv\int\xi[{\bar
u}(\xi)+{\bar d}(\xi)]d\xi$ and $S^+\equiv\int\xi[s(\xi)+{\bar
s}(\xi)]d\xi$. With a calculation of the $Q_V$, $Q_S$ and $S^+$
from CTEQ5 parametrization at $Q^2=16$GeV$^2$, together with
$|V_{cd}|^2=0.05$ and $|V_{cs}|^2=0.95$, the ratio $R$ is
estimated to be about $20\%$ ($25\%$) for $2S^-/Q_V$ being 0.007
(0.022) from Table~1. Thus the cross section difference
$f_{H^+}-f_{H^-}$ is a significant quantity that can be extracted
from the semi-inclusive differential cross sections.

Neutrino experiment with emulsion target, like the CHORUS
detector, is ideal for the study of charmed hadron production.
Compared to dimuon studies, it has a much lower level of
background and is free from the uncertainties that exist in charm
muonic weak decay processes~\cite{CHOR02}. And for statistics,
CHORUS reported in total about 94000 neutrino CC events located
and fully reconstructed, in which about 2000 charm events were
observed~\cite{CHOR04,lel04}. This has been compatible with dimuon
statistics. If such (or higher) statistics can be achieved with
both neutrino and antineutrino beams of high energies in future
experiments, the question about strange asymmetry is promising to
be settled.

\section{Light quark fragmentation}

The possibility that a light quark fragments into charmed hadrons
(associated charm production) can be an interesting effect of
non-perturbative QCD, and it has been explored~\cite{god8489} to
explain the unexpected high rate of like-sign dimuons production
from many neutrino experiments~\cite{CDHSW,CCFR93LSD,CHARM}.
Although the field has been inactive for years, and in practice
people generally assume the light quark fragmentation (LQF) to be
negligible, the physical possibility of a small contribution is
not ruled out. In fact, as to our consideration, neutrino
experiments can be slightly different from $e^{+}e^{-}$
experiments in this respect. The scattered light quark with high
momentum can pick up a charm quark or antiquark from nucleon sea
to form a $D$ meson, and the larger the energy of the light quark,
the more the ability that it can pick up a charm from the sea.
This energy dependence is apparent in prompt like-sign dimuons
production rates in many experiments~\cite{CDHSW,CCFR93LSD,CHARM}.
Since the scattered quark has most of the energy in the collision,
it is much more promising to pick up the charm quark from nucleon
sea than other produced quarks in fragmentation process. Thus as
to our consideration such fragmentation as $u\rightarrow {\bar
D}^{0}(\bar{c}u)$, $d\rightarrow{D^{-}}(\bar{c}d)$ can possibly be
non-negligible in high energy neutrino experiments.


In case that light quark can fragment into charmed hadrons, the
process can manifest itself in a number of observables, such as
the prompt like-sign dilepton and trimuon productions in high
energy neutrino experiments~\cite{smi83}, direct observations of
two charmed hadrons in nuclear emulsion target~\cite{CHOR02ONE},
and charm production in hadron collisions~\cite{E791D96,dia00}.
Among these, the prompt like-sign dimuon productions have been the
most seriously studied and we will use some of the data for a
quantitative estimate of light quark fragmentation function and
its influence on the extraction of strange asymmetry from CC charm
production processes.

Prompt like-sigh dimuons (${\mu^-}{\mu^-}$) can be produced
through the process: $\nu+d\rightarrow\mu^-+u$, with the scattered
$u$ to fragment into ${\bar D}^0 ({\bar c}u)$ or ${\bar
D}^{*0}({\bar c}u)$ meson. Its differential cross section can be
expressed as:
\begin{eqnarray}
{d^3\sigma_{\nu N\rightarrow\mu^-\mu^-X}\over dx dy
dz}={2G^2ME_\nu|V_{ud}|^2\over \pi(1+Q^2/M_W^2)^2} x
{u(x)+d(x)\over 2}D_q(z)B_{{\bar D}^0}\;,\label{nn}
\end{eqnarray}
where $D_q(z)$ is the total fragmentation function for a light
quark to fragment into charmed hadrons, defined as $D_q(z)\equiv
D_q^D(z) +D_q^{D^*}(z)$, with $D_q^D(z)\equiv D_u^{\bar
D^0}(z)=D_{\bar u}^{D^0}(z)=D_d^{D^-}(z)=D_{\bar d}^{D^+}(z)$ and
$D_q^{D^*}(z)\equiv D_u^{\bar D^{*0}}(z)=D_{\bar
u}^{D^{*0}}(z)=D_d^{D^{*-}}(z)=D_{\bar d}^{D^{*+}}(z)$ simply
assumed. Note here that $D_q(z)$ is energy dependent in analogous
to containing an energy suppression factor for charm production.
$B_{{\bar D}^0}$ is the inclusive muonic decay ratio for ${\bar
D}^0$ meson decay ${\bar D}^0\rightarrow\mu^-X$, which is the same
for ${\bar D}^{*0}$ meson, since all ${\bar D}^{*0}$ will decay
into ${\bar D}^{0}$ at first.

Similarly, the differential cross section for prompt ${\mu^+}
{\mu^+}$ production in $\bar\nu$ induced DIS on isoscalar target
is
\begin{eqnarray}
{d^3\sigma_{\bar\nu N\rightarrow\mu^+\mu^+X}\over
dxdydz}={2G^2ME_\nu|V_{ud}|^2\over \pi(1+Q^2/M_W^2)^2} x {\bar
u(x)+\bar d(x)\over 2}D_q(z)B_{D^0}\;.\label{pp}
\end{eqnarray}

Many experimental groups have reported positive results on prompt
like-sign dimoun production. Among them the CDHSW~\cite{CDHSW} and
CCFR~\cite{CCFR93LSD} data have a high precision and show much
consistency with each other. Another high precision experiment
CHARM~\cite{CHARM}, which has reported a much higher
$\mu^{-}\mu^{-}$ production rate, received doubts on their
estimate of $\pi/K$ decay background~\cite{CDHSW}. Besides, their
kinematic cut $p_{\mu}>4$ GeV, which is lower than other
experiments ($p_{\mu}>9$ GeV), can permit more $\mu^{-}\mu^{-}$
events and thus can produce a higher rate. Since kinematic cut on
the second $\mu$ reduces event number and thus the rate of
like-sign dimuons in CC events: $\mu^{-}\mu^{-}$/$\mu^{-}$, it
will probably underestimate the light quark fragmentation (LQF)
effect to use the $\mu^{-}\mu^{-}$/$\mu^{-}$ data. On the other
hand, the ratio of prompt like-sign dimuons to opposite-sign
dimuons $\mu^{-}\mu^{-}$/$\mu^{-}\mu^{+}$ is expected to be less
influenced by kinematic cut, as both second muons receive the same
kinematic cut. Thus, we consider it appropriate to use the
$\mu^{-}\mu^{-}$/$\mu^{-}\mu^{+}$ data other than
$\mu^{-}\mu^{-}$/$\mu^{-}$ data to estimate the LQF effect.

The differential cross section for $\mu^{-}\mu^{+}$ production in
$\nu$ induced CC DIS on an isoscalar target is given by
\begin{eqnarray}%
{d^3\sigma_{\nu N\rightarrow\mu^-\mu^+X}\over d\xi dydz}
&=&{2G^2ME_\nu\over \pi(1+Q^2/M_W^2)^2}f_c\bar B_c(z)\nonumber\\
&\times&\left[\xi{d(\xi)+u(\xi)\over 2}|V_{cd}|^2 + \xi
s(\xi)|V_{cs}|^2\right] \nonumber\\
 &+&\delta\left({d^3 \sigma_{\nu
N\rightarrow\mu^-\mu^+X}\over d\xi dydz}\right)_{\mathrm{LQF}}
\;,\label{np}
\end{eqnarray}
where $f_c$ is the charm suppression factor $f_c\equiv
1-{m_c^{2}}/{2ME_{\nu}\xi}$, and $\bar B_c(z)$ is the average
muonic decay ratio for the charmed hadrons produced in CC DIS,
$\bar B_c(z)=\sum_H D_c^H(z) B_H$, with H being $D^0$, $D^{*0}$,
$D^+$, $D^{*+}\cdots$. The last term marked with LQF is the light
quark fragmentation contribution to $\mu^{-}\mu^{+}$ production
($\nu\bar{u}\rightarrow\mu^{-}\bar{d}\rightarrow
\mu^{-}D^{+}(D^{*+})X'\rightarrow\mu^{-}\mu^+X$),
\begin{eqnarray}%
\delta\left({d^3\sigma_{{\nu}N\rightarrow\mu^-\mu^+ X}\over d\xi
dy dz}\right)_{\mathrm{LQF}} &=&{2G^2ME_\nu|V_{ud}|^2\over
\pi(1+Q^2/M_W^2)^2}D_q(z)\bar B_{D^{(*)+}}\nonumber\\
 &\times& \left[x{\bar d(x)+\bar
u(x)\over 2}\right](1-y)^2\;,~~~\label{nplqf}
\end{eqnarray}
where $\bar B_{D^{(*)+}}$ is the average muonic decay ratio of the
$D^+$ and $D^{*+}$ mesons produced from $\bar d$ quark
fragmentation, $\bar B_{D^{(*)+}}={1\over D_q}(D_q^D B_{D^+}
+D_q^{D^*} B_{D^{*+}})$. Since $D^{*+}$ decays to $D^0$ with
branching ratio $B\simeq67.7\%$~\cite{PDG04}, and to $D^+$ with
ratio $1-B$, we have $\bar B_{D^{(*)+}}=b B
B_{D^0}+(1-bB)B_{D^+}$, with $b\equiv D_q^{D^*}/D_q$.

Similarly, the differential cross section for $\mu^{+}\mu^{-}$
production in $\bar\nu$ induced CC DIS on isoscalar target reads
\begin{eqnarray}%
{d^3\sigma_{\bar\nu N\rightarrow\mu^+\mu^-X}\over d\xi dydz}
&=&{2G^2ME_\nu\over \pi(1+Q^2/M_W^2)^2}f_c\bar B_{\bar c}(z) \nonumber\\
&\times&\left[\xi{\bar d(\xi)+\bar u(\xi)\over 2}|V_{cd}|^2 + \xi
\bar s(\xi)|V_{cs}|^2\right] \nonumber\\
 &+&\delta\left({d^3 \sigma_{\bar\nu
N\rightarrow\mu^+\mu^-X}\over d\xi dydz}\right)_{\mathrm{LQF}}
\;,\label{pn}
\end{eqnarray}
with
\begin{eqnarray}%
\delta\left({d^3\sigma_{{\bar\nu}N\rightarrow\mu^+\mu^- X}\over
d\xi dy dz}\right)_{\mathrm{LQF}} &=&{2G^2ME_\nu|V_{ud}|^2\over
\pi(1+Q^2/M_W^2)^2} D_q(z)\bar B_{D^{(*)-}} \nonumber\\
&\times&\left[x{d(x)+ u(x)\over
2}\right](1-y)^2\;.~~~~~\label{pnlqf}
\end{eqnarray}

CDHSW has reported prompt dimuon rates
$\sigma_{\mu^{-}\mu^{-}}$/$\sigma_{\mu^{-}\mu^{+}}$ and
$\sigma_{\mu^{+}\mu^{+}}$/$\sigma_{\mu^{+}\mu^{-}}$ in $\nu$ and
$\bar\nu$ induced DIS. As mentioned previously, these data are
less influenced by kinematic cut and thus are better suited for
the extraction of the LQF effect. The prompt dimoun rates from
CDHSW with visible energy $E_{\mathrm{vis}}$ in range $100\sim200$
GeV are listed in Table~2.
\begin{table*}
\caption{Prompt dimuon rates for $100<E_{\mathrm{vis}}<200$
GeV~\cite{CDHSW}}
\label{tab:2}       
\begin{tabular}{lllll}
\hline\noalign{\smallskip}
  &
$\sigma_{\mu^{-}\mu^{-}}/\sigma_{\mu^{-}\mu^{+}}$
         &$\sigma_{\mu^{-}\mu^{-}}/\sigma_{\mu^{-}}$
         &$\sigma_{\mu^{+}\mu^{+}}/\sigma_{\mu^{+}\mu^{-}}$
         & $\sigma_{\mu^{+}\mu^{+}}/\sigma_{\mu^{+}}$ \\
          \noalign{\smallskip}\hline\noalign{\smallskip}
$p_{\mu} > 6$ GeV  & $(3.5\pm 1.6)\%$ & $(1.6\pm
0.74)\times10^{-4}$&
 $(4.5\pm 2.0)\%$ & $(2.2\pm 1.0) \times 10^{-4}$ \\
$p_{\mu} > 9$ GeV &
 $(2.9\pm1.2)\%$  & $(1.05\pm 0.43)\times 10^{-4}$&
 $(4.4\pm 1.8)\%$ & $(1.7\pm 0.7) \times 10^{-4}$ \\
$p_{\mu} > 15$ GeV &
 $(2.3\pm 1.0)$\% & $(0.52\pm 0.22)\times 10^{-4}$ &
$(4.1\pm 2.3)\%$  & $(0.8\pm 0.45)\times 10^{-4}$
\\
\noalign{\smallskip}\hline
\end{tabular}
\end{table*}

As can be seen from Table 2, the prompt dimuons rate
$\sigma_{\mu^{-}\mu^{-}}$/$\sigma_{\mu^{-}\mu^{+}}$ still show a
slight dependence on kinematic cut, though much smaller than the
$\sigma_{\mu^{-}\mu^{-}}$/$\sigma_{\mu^{-}}$ data do. Thus we can
only estimate the order of magnitude for the LQF effect with the
reported data.

The rate $\sigma_{\mu^{-}\mu^{-}}$/$\sigma_{\mu^{-}\mu^{+}}$
(without any kinematic cut) can be deduced by Eq.~(\ref{nn}) and
Eq.~({\ref{np}) with an integral on kinematic variables to
approximate
\begin{eqnarray}
{\sigma_{\mu^{-}\mu^{-}}\over
\sigma_{\mu^{-}\mu^{+}}}\approx{Q_{ud}|V_{ud}|^2\over
Q_{ud}|V_{cd}|^2+S|V_{cs}|^2} \cdot {D_qB_{D^0}\over \bar f_c \bar
B_c}\;,
\end{eqnarray}
where $Q_{ud}\equiv {1\over 2}\int x[u(x)+d(x)]dx$, $S\equiv\int
xs(x)dx$,  and $\bar f_c$ denotes the average of energy
suppression factor in Eq.~(\ref{np}). With the measured
$B_{D^0}\simeq 6.87\%$ and $\bar B_c\simeq 8.8\%$~\cite{lel04},
together with $Q_{ud}$ and $S$ from CTEQ5 at $Q^2=16$ GeV$^2$,
$D_q/\bar f_c$ is estimated to be:
\begin{eqnarray}
{D_q\over\bar f_c}\approx 0.199 {\sigma_{\mu^{-}\mu^{-}}\over
\sigma_{\mu^{-}\mu^{+}}}\;.\label{Dq}
\end{eqnarray}
With the experimental data on ${\sigma_{\mu^{-}\mu^{-}}/
\sigma_{\mu^{-}\mu^{+}}}$ from Table~2, one can easily estimate
$D_q/\bar f_c$ by Eq.~(\ref{Dq}).

Since we are most interested in the strange quark-antiquark
asymmetry here, we will directly address the influence of LQF
effect on the extraction of strange asymmetry. Because LQF effect
contributes differently for ${\nu}$ induced $\mu^{-}\mu^{+}$
production and for $\bar{\nu}$ induced $\mu^{+}\mu^{-}$
production, it will give different corrections to $s$ and
$\bar{s}$ distributions, and thus influence the measurement of
strange asymmetry from opposite-sign dimuon method. To illustrate
this, we will compare the contribution of LQF effect with that of
strange asymmetry on the difference between $\nu$  and $\bar{\nu}$
induced opposite-sign dimuon production cross sections. The latter
(strange asymmetry contribution) can be drawn from model
predictions in the last column of Table 1, when assuming the
average muonic branding ratio of charmed hadrons to be the same
for $\nu$ and $\bar{\nu}$ induced CC DIS $\bar B_c(z)=\bar B_{\bar
c}(z)$. The former (LQF contribution) can be deduced from
Eq.~({\ref{np})-Eq.~({\ref{pnlqf}) with an assumption $\bar
B_{D^{(*)+}}=\bar B_{D^{(*)-}}$, and be compared to the strange
asymmetry part with an integral on kinematic variables. The
fraction of the LQF contribution is
\begin{eqnarray}%
P_{\mathrm{LQF}}&\equiv&{\delta(\sigma_{{\nu}N\rightarrow\mu^-\mu^+
X}-\sigma_{{\bar\nu}N\rightarrow\mu^+\mu^- X})_{\mathrm{LQF}}\over
(\sigma_{{\nu}N\rightarrow\mu^-\mu^+
X}-\sigma_{{\bar\nu}N\rightarrow\mu^+\mu^- X})_{\mathrm{total}}}
\nonumber
\\&\approx& -{{1\over 3}Q_V|V_{ud}|^2\over
Q_V|V_{cd}|^2+2S^-|V_{cs}|^2} \cdot {D_q\bar B_{D^{(*)+}}\over
\bar f_c \bar B_c}\;.\label{PLQF}
\end{eqnarray}
To assess $P_{\mathrm{LQF}}$, the value of $\bar B_{D^{(*)+}}$ is
needed. Remember that $\bar B_{D^{(*)+}}=b B
B_{D^0}+(1-bB)B_{D^+}$, with $b\equiv D_q^{D^*}/D_q$. The unknown
$b$ is the fraction of vector $D^{*}$ meson in light quark
fragmentation. When we set $b$ to be ${{1}/{3}}\sim{{2}/{3}}$, and
take $B_{D^0}\simeq6.87\%$, $B_{D^+}\simeq17.2\%$~\cite{lel04}, we
get $\bar B_{D^{(*)+}}=(13.7\pm1.2)\%$. Using Eq.~(\ref{PLQF}) and
taking $2S^{-}/Q_V=0.007$ from Table~1, we get
$P_{\mathrm{LQF}}=-(1.73\pm0.15){\sigma_{\mu^{-}\mu^{-}}}/{\sigma_{\mu^{-}\mu^{+}}}$.
Taking
${\sigma_{\mu^{-}\mu^{-}}}/{\sigma_{\mu^{-}\mu^{+}}}=(3.5\pm1.6)\%$
from Table~2, we get
\begin{eqnarray}
P_{\mathrm{LQF}}=-(6.1^{+3.5}_{-3.1})\%\,.
\end{eqnarray}
Thus, we get an estimate of the LQF contribution to be a few
percent compared to strange asymmetry contribution
$P_{\mathrm{SA}}:12\%\sim40\%$. However, the constraint of
$P_{\mathrm{LQF}}$ can also be done with
${\sigma_{\mu^{+}\mu^{+}}}/{\sigma_{\mu^{+}\mu^{-}}}$ data, and
the result is $P^\prime_{\mathrm{LQF}}=-(33^{+19}_{-16})\%$, which
is very large compared to result from the
${\sigma_{\mu^{-}\mu^{-}}}/{\sigma_{\mu^{-}\mu^{+}}}$ data. This
large discrepancy is difficult to explain at present, and may
imply an uncertainty in the estimate of the LQF contribution in
the opposite-sign dimuon measurements of strange asymmetry.

From the sign and size of $P_{\mathrm{LQF}}$, one sees that the
LQF effect contributes oppositely to the predicted strange
asymmetry contribution on the whole, with a rate that could be
non-negligible in opposite-sign dimuon experiments.


The LQF effect also exists in the process of inclusive charm
productions that we suggest. For $D^{\pm}$ production, the cross
section difference, $f_{D^{+}}-f_{D^{-}}$, for $\nu$ and
$\bar{\nu}$ induced CC DIS will include an additional term from
light quark fragmentation:
\begin{eqnarray}
\delta (f_{D^+}-f_{D^-})_{\mathrm{LQF}}
&=&-{2G^2ME_\nu|V_{ud}|^2\over
\pi(1+Q^2/M_W^2)^2}D_q(z)(1-\varepsilon)\nonumber\\
&\times& \left[x {d_v(x)+u_v(x)\over
2}\right](1-y)^2\;,~~~~~\label{D+LQF}
\end{eqnarray}
where $\varepsilon=Bb$ is introduced with the consideration that
part of $D^{*+}(D^{*-})$ will decay into $D^{0}(\bar D^{0})$ and
will not contribute to the cross sections.

For neutral charm production, LQF contributes to $\bar D^0$
production in $\nu$ induced CC DIS ($\nu+d\rightarrow\mu^-+u$,
$u\rightarrow\bar D^0(\bar{c}u)$), and to $D^0$ production in
$\bar{\nu}$ induced CC DIS. In case that $D^0$ and $\bar D^0$ are
not distinguished by emulsion target, the $\bar D^0$ $(D^0)$
production in $\nu$ ($\bar{\nu}$) induced CC DIS from LQF will be
incorporated to $D^0$ $(\bar D^0)$ production in $\nu$
($\bar{\nu}$) induced CC DIS. Thus an additional term from LQF
will contribute to $f_{D^0}-f_{\bar D^0}$:
\begin{eqnarray}
\delta (f_{D^0}-f_{\bar D^0})_{\mathrm{LQF}}
&=&{2G^2ME_\nu|V_{ud}|^2\over
\pi(1+Q^2/M_W^2)^2}D_q(z)(1-\varepsilon')\nonumber\\
&\times& \left[ x {d_v(x)+u_v(x)\over 2}\right]\;,\label{D0LQF}
\end{eqnarray}
where $\varepsilon'=(1-y)^2Bb$, which is introduced from $\bar d$
($d$) fragmentation into $D^{*+}$ $(D^{*-})$ mesons that then
decay into $D^0$ $(\bar D^0)$ and contribute to cross section
difference $f_{D^0}-f_{\bar D^0}$.


The proportion of LQF contribution to inclusive charm production
cross section difference $f_{H^+}-f_{H^-}$, namely
$P^{H^{\pm}}_{\mathrm{LQF}}$, can be estimated similarly to that
of dimuon productions. With an integral on kinematic variables of
Eq.~(\ref{csd}), Eq.~(\ref{D+LQF}), Eq.~(\ref{D0LQF}), and using
charm production fractions $\int D_c^{D^+}(z)dz\simeq 0.26$ and
$\int D_c^{D^0}(z)dz\simeq 0.66$ for $E_{\nu}>80$GeV~\cite{lel04},
$P^{H^{\pm}}_{\mathrm{LQF}}$ is estimated (in unite of
$P_{\mathrm{LQF}}$) to be: $P^{D^{\pm}}_{\mathrm{LQF}}\approx
1.6P_{\mathrm{LQF}}$ for $D^{\pm}$ meson productions, and
$P^{D^{0}}_{\mathrm{LQF}}\approx -2.6P_{\mathrm{LQF}}$ for $D^0$,
$\bar D^0$ meson productions.

If the LQF contribution $P_{\mathrm{LQF}}$ in opposite-sign
dimuons measurement is in the order of a few percent percent and
opposite to strange asymmetry contribution $P_{\mathrm{SA}}:
12\%\sim 40\%$, just as we have estimated, the LQF will contribute
to inclusive charm production with a larger proportion (in the
order of about ten percent or even larger). For inclusive
$D^{\pm}$ production, LQF contributes oppositely compared to
strange asymmetry when $x[s(x)-\bar s(x)]>0$. On the other hand,
for inclusive CC neutral charm $(D^0, \bar D^0)$ production, LQF
contributes positively compared to strange asymmetry when
$x[s(x)-\bar s(x)]>0$. A separation of the LQF effect and the
strange asymmetry effect can be made from the distinct features of
$f_{D^+}-f_{D^-}$ and $f_{D^0}-f_{\bar D^0}$ measured by nuclear
emulsion target. Thus, the inclusive measurement of charged and
neutral charm production in $\nu$ and $\bar{\nu}$ induced CC DIS
will shed light on both the strange asymmetry and the LQF effect.

Dedicated analysis of charm productions in neutrino experiments
and in other processes will be helpful for a more precise estimate
and constraint for the light quark fragmentation effect.

\section{Conclusions}

For probing the nucleon strange asymmetry, we analyzed the charged
current charm production processes, in particular, the $\nu_\mu$
induced $H^+$ ($D^+$ or $D^0$) production and the $\bar\nu_\mu$
induced $H^-$ ($D^-$ or $\bar D^0$) production processes. The
strange asymmetry from various model calculations that can explain
the NuTeV anomaly is shown in general to contribute a sizable
proportion ($12\%\sim 40\%$) to the $H^\pm$ differential cross
section difference $f_{H^+}-f_{H^-}$. Thus, measurement of these
cross sections with high energy neutrino and antineutrino beams on
nuclear emulsion target is very promising to detect the strange
quark-antiquark asymmetry.

Meanwhile, we analyzed the possible light quark fragmentation
(LQF) effect from prompt like-sign dimuon data and studied its
influence on the measurement of strange asymmetry. Our result is
that the LQF may be an important source that reduces the effect of
strange asymmetry from opposite sign dimuon studies. And for
inclusive charged current (CC) charm production with emulsion
target, since the contributions of LQF are in opposite directions
for $D^{\pm}$ and for $D^{0}$ ($\bar D^{0}$) productions, a
separation of the LQF effect from strange asymmetry effect can be
made by the separate measurement of $D^{\pm}$ and neutral charm
differential cross sections in CC DIS. Thus the inclusive
measurement of charmed hadrons can shed light on both strange
asymmetry and the LQF effect. Further analysis and constraint for
LQF effect from various experiments will also be helpful for the
purpose of measuring the strange asymmetry more reliably.

{\bf Acknowledgments: } We acknowledge the helpful discussion with
Vincenzo Barone. This work is partially supported by National
Natural Science Foundation of China (Nos.~10025523, 90103007, and
10421003), by the Key Grant Project of Chinese Ministry of
Education (No.~305001), and by the Italian Ministry of Education,
University and Research (MIUR).


\end{document}